\documentclass[epsf]{kluwer}
\usepackage{epsf}

\def\kms{km~s$^{-1}$}

\begin{document}
\begin{article}

\begin{opening}

\title{What is the behavior of the ISM in the SMC?}

\author{Snezana \surname{Stanimirovic}\email{sstanimi@naic.edu}} 
\institute{Arecibo Observatory, NAIC/Cornell University, HC3 Box 53995,
Arecibo, Puerto Rico 00612}
\runningtitle{What is the behavior of the ISM in the SMC}
\runningauthor{S. Stanimirovic}
 
\begin{abstract}
We describe quantitatively the neutral hydrogen (HI) and dust content
of the interstellar medium (ISM) in the Small Magellanic Cloud (SMC),
using the spatial power spectrum. The velocity modification of the HI
density power spectrum is investigated and discussed.    
\end{abstract}
\end{opening}

\section{Introduction}
Many observations in the past decade have challenged the traditional
picture of the interstellar medium (ISM). Instead of  a two-level
hierarchical system, consisting of clouds uniformly dispersed in the
intercloud medium, the ISM shows an astonishing inhomogeneity, with many
levels of hierarchy.  In order to consider real density functions in
physical processes, a better understanding of the inventory and topology of
the ISM is essential, as well as the processes responsible for their
creation.  Having an extremely gas-rich ISM, dwarf irregular galaxies are
particularly suitable for such studies. We hence here describe
quantitatively, using the spatial power spectrum, the inventory of the  ISM
in the Small Magellanic Cloud\footnote{We assume SMC to be at the distance
of 60 kpc throughout  this study.} (SMC) and point to several processes
that may be involved  in the sculpturing of its ISM.

\section{HI and IR spatial power spectrum}

\begin{figure}
\begin{center}
\leavevmode
\epsfysize=5cm
\epsffile{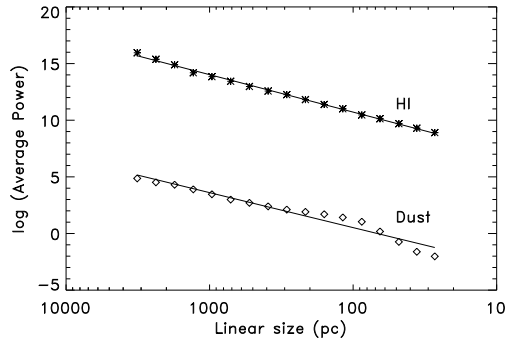}
\end{center}
\caption{\label{f:power-spec}The 2-D spatial power spectra of HI
(asterisks) and dust (diamonds) mass column densities (in units of M$_{\odot}$).}
\end{figure}

The power spectrum of the HI emission fluctuations in the SMC was derived
in Stanimirovic et al. (1999). For more information on the exact technique
and the data used see \inlinecite{Stanimirovic99}.  It was shown that the
2-D power spectra can be remarkably well fitted by a power law, $P(k)
\propto k^{\gamma}$, over the continuous  range of spatial scales $\sim$ 30
pc -- 4 kpc, and over the velocity range 110 -- 200 \kms. Using the
velocity slices of $\sim$ 20 \kms, the average slope was estimated with
$\langle \gamma \rangle=-3.04\pm0.02$. No change of the power law slope
was seen on either large or small scale end. However, when looking at the
power spectrum of the HI column density distribution, after integrating
along the whole velocity range, a change of the power law slope  was
noticed, with $\gamma$ being equal $-3.31\pm0.01$ \cite{Stanimirovic00}.
The power spectrum of dust column density fluctuations in the SMC was
derived in \inlinecite{Stanimirovic00}. This spectrum can be fitted by
$P_{\rm d}(k) \propto k^{-3.1\pm0.2}$. A slight change  of slope on spatial
scales smaller than 50 pc may be present though.  Nevertheless, slopes for
HI and dust column density power spectra appear  to be very similar, see
Fig.~\ref{f:power-spec}.

The power law fit of both HI and dust column-density power spectra shows
that the hierarchical structure organization is present within both HI and
dust content of the ISM in the SMC, with no preferred spatial scales for
both HI and dust clouds. Similar power law indices suggest that similar
processes are involved in shaping both HI and dust content of the SMC.

\subsection{Velocity modification of the HI power spectrum}

The 2-D intensity fluctuations traced by the power spectra have
contribution from both density and velocity fluctuations. Indeed, due to
the velocity fluctuations, two clumps along the same line of sight at
different distances may appear in the same velocity channel, hence
doubling the measured intensity. It is therefore necessary to disentangle
density from velocity influence to the power spectrum.  The importance of
this phenomenon was first recognized by \inlinecite{Lazarian99}. They start
with 3-D density spectrum in velocity space ($P_{s} \propto K^{n}$) and
calculate, analytically and numerically, 2-D power spectrum of intensity
fluctuations, in two particular cases: (a) the 3-D density spectrum is
small-scale dominated ($n>-3$); and (b) the 3-D density spectrum is
large-scale dominated ($n<-3$). One of the main results in
\inlinecite{Lazarian99}  is that the intensity statistics depends strongly
on velocity slice thickness.

\begin{figure}
\begin{center}
\leavevmode
\epsfysize=5cm
\epsffile{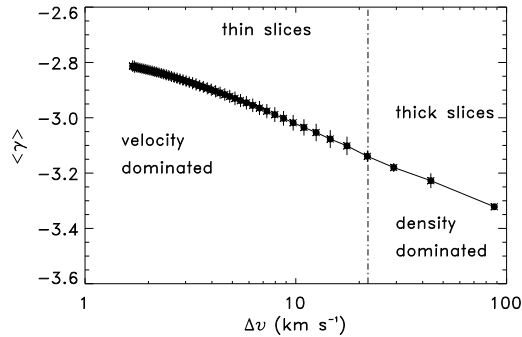}
\end{center}
\caption{\label{f:gamma-variation} The variation of the 2-D HI power spectrum
slope $\langle \gamma \rangle$ with the velocity slice thickness $\Delta
v$. The dot-dashed line distinguishes thin from thick slice regimes. }
\end{figure}

To test the predictions by \inlinecite{Lazarian99} in the case of the SMC,
we have determined the power spectrum slope, $\langle \gamma \rangle$,
while varying the velocity slice thickness from $\sim$ 2 \kms~to $\sim$
100 \kms. A significant variation of $\langle \gamma \rangle$ was found,
shown in Fig.~\ref{f:gamma-variation}, consistent with the predictions ---
$\langle \gamma \rangle$ decreases with an increase of velocity slice
thickness. The thickest velocity slice gives $n=-3.3$ suggesting that we
are in the large-scale dominated regime. Hence, the intensity spectrum is
dominated by velocity fluctuations and only the thickest velocity slices
must be used in order to find density fluctuations. Using the thin slices,
however, we can find the slope of velocity fluctuations to be $m=0.4$. The
transition point between thin and thick slice regimes is equal to the
velocity dispersion on the scale of the whole SMC ($\sim$ 4 kpc), which is
$\sim$ 22 \kms.  Both $n$ and $m$ are significantly shallower than for the
case of  Kolmogorov turbulence (where $n=-11/3$ and $m=2/3$).

\section{On the origin of the power spectrum}

The hierarchical structure organization is usually ascribed to interstellar
turbulence (see \opencite{Elmegreen00}). But which processes create this
turbulence is still not well understood. However, there are several
possible candidates. As a large number of expanding shells was found in the
SMC, the energy injection by  these shells can significantly stir up the
ISM. The radii of the SMC shells range from $\sim$ 30 pc to $\sim$ 2 kpc,
with most of them being around 100 pc. However, no specific scales on which
the energy injection happens show up in the HI power spectra.  Very
recently \inlinecite{Goldman00} suggested a very different scenario:  large
scale turbulence is induced by instabilities in the large-scale flows
during the last SMC--LMC encounter. In this case both dust and gas are
just `passive markers', they do follow turbulent field but do not feed back
dynamically.  \inlinecite{Elmegreen00} shows that interacting, nonlinear
magnetic waves can produce hierarchical density structure out of an
initially uniform medium. Actually, the power spectrum of such simulated
structure has a power law slope between $-2.5$ and $-3.6$, which is close
to what was observed.

\section{Summary}
The spatial power spectrum of HI and dust content in the SMC is well fitted
by a power law, with power law slopes being similar.  The HI spectrum
appears to be modified by velocity fluctuations. After disentangling
velocity from density fluctuations, the  3-D HI density spectrum has slope
of $-3.3$. This is significantly shallower than for Kolmogorov turbulence.

\begin{acknowledgements}
The stimulating discussions with Alex Lazarian, Dmitri Pogosyan and Steve
Shore are greatly appreciated. 
\end{acknowledgements}

\end{article}
\end{document}